\documentclass[aps, prd ]{revtex4}

\usepackage{amsmath,amsfonts,amscd,amssymb,epsf}
\usepackage[small]{caption}

\newcommand{\pa}{\partial}

\begin{document}

 \title{The Casimir interaction of a massive vector field between concentric spherical bodies}
\author{L. P. Teo}\email{ LeePeng.Teo@nottingham.edu.my}\address{Department of Applied Mathematics, Faculty of Engineering, University of Nottingham Malaysia Campus, Jalan Broga, 43500, Semenyih, Selangor Darul Ehsan, Malysia. }
\begin{abstract}
The Casimir interaction energy  due to the vacuum fluctuations of a massive vector field between two perfectly conducting concentric spherical bodies is computed.  The  TE contribution to the Casimir interaction energy is a direct generalization of the massless case but the TM  contribution is much more complicated. Each TM mode is a linear combination of a transverse mode which is the generalization  of a TM mode in the massless case and a longitudinal mode that does not appear in the massless case.  In contrast to the case of two parallel perfectly conducting plates, the are no TM discrete modes that vanish identically in the perfectly conducting spherical bodies. Numerical simulations show that the Casimir interaction force between the two bodies is always attractive.
\end{abstract}
\maketitle

\section{Introduction}
Casimir effect has been under intensive investigation for its   relation to many other areas of physics \cite{11}. It has been studied for both massless and massive scalar fields as well as spinor fields in different geometric configurations. For electromagnetic fields (massless vector fields), the situation is more subtle. Although Casimir effect of electromagnetic fields has been extensively studied,  there are very few works that considered Casimir effect of massive vector fields. The pioneering work of Barton and Dombey \cite{1} considered Casimir effect of a massive vector field on a pair of perfectly conducting plates in vacuum. This work revealed   that the analysis for a massive vector field is much more complicated   than for a massless vector field. One of the fundamental difference is that for a massless vector field, one can regard the electric field and magnetic field as the primary quantities and there is a gauge degree of freedom in the potentials. However, for a massive vector field, the primary quantities are the scalar and the vector potentials and the gauge freedom is lost. A massless vector field cannot penetrate through a perfectly conducting plate but for a massive vector field, although the electric field and the magnetic field have to vanish in a perfectly conducting plate, the potentials need not vanish. The latter implies that even for the Casimir effect of a massive vector field on a pair of perfectly conducting plates, one has to treat it as if considering Casimir effect in dielectric plates. In fact, it was shown in \cite{1} that in the configuration of two parallel perfectly conducting plates, the field modes can be divided into three types. Two of these are discrete modes where the potentials vanish identically in the plates, and the third type is continuum modes with nonvanishing potentials in the plates. The formula for the contribution of the continuum modes to the Casimir energy can be regarded as a special case of the Lifshitz formula \cite{11,10,7} for the Casimir energy between dielectric plates.

In \cite{2}, we have generalized the work \cite{1} and considered the Casimir effect on a pair of magnetodielectric plates in an arbitrary medium.  We found that   the Casimir effect of a massive vector field and a massless vector field   have significant differences. As a result, it is important to investigate the Casimir effect of massive vector fields in geometric configurations other than the parallel plates.
In this article, we consider the Casimir effect of a massive vector field  between two concentric perfectly conducting spherical bodies. The Casimir effect of a massless vector field on   two perfectly conducting spherical shells   have been considered in \cite{3,4,6,5,12}. It was shown that the Casimir force always tends to attract the two shells to each other.  It will be interesting to see whether this is still true for a massive vector field.

\section{Casimir energy of two concentric perfectly conducting spherical bodies  }
\begin{figure}[h]
\epsfxsize=0.35\linewidth \epsffile{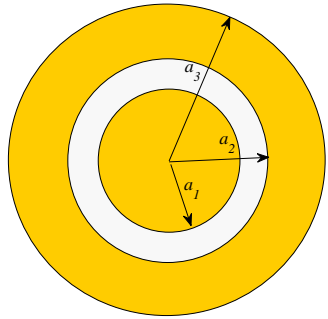} \caption{\label{f1} Two concentric perfectly conducting spherical bodies. }\end{figure}
In this section, we compute the Casimir energy between two concentric perfectly conducting spherical bodies (see Fig. \ref{f1}) due to the vacuum fluctuations of a massive vector field. A massive vector field is represented by a  four-vector $\displaystyle\left\langle \frac{\varphi}{c}, A_x, A_y, A_z\right\rangle$, where $\varphi$ is the scalar potential and $\mathbf{A}=\langle A_x, A_y, A_z\rangle$ is the vector potential. The electric field $\mathbf{E}$ and the magnetic field $\mathbf{B}$ are given respectively by \begin{align}\label{eq1_5}\mathbf{E}=-\frac{\pa\mathbf{A}}{\pa t}-\nabla\varphi,\hspace{1cm}\mathbf{B}=\nabla\times \mathbf{A}.\end{align}
From these, one immediately obtain two of the Maxwell's equations:
\begin{equation}\label{eq12_1_1}
\nabla\cdot\mathbf{B}=0, \quad\nabla\times\mathbf{E}+\frac{\pa\mathbf{B}}{\pa t}=0.
\end{equation}
For a massive vector field, the other two Maxwell's equations have to be modified. In a magnetodielectric medium with    permittivity $\varepsilon$ and permeability $\mu$, assume that the usual linear relations $\mathbf{D}=\varepsilon \mathbf{E}$ and $\mathbf{B}=\mu\mathbf{H}$ hold. Then
\begin{equation}\label{eq11_15_1}\begin{split}
&\nabla\cdot\mathbf{D}+\frac{m^2 }{\mu\hbar^2}\varphi=  \rho_f,  \\
&\nabla\times \mathbf{H}-\frac{\pa\mathbf{D}}{\pa t}+ \frac{m^2c^2}{ \mu\hbar^2}\mathbf{A}=\mathbf{J}_f,
\end{split}\end{equation}where $\rho_f$ and $\mathbf{J}_f$ are the free charges and the free current in the medium. Eqs.
\eqref{eq12_1_1} and \eqref{eq11_15_1} together constitute the Proca equations for a massive vector field \cite{1}.
The potentials $\varphi$ and $\mathbf{A}$ satisfy the Lorentz condition
\begin{equation}\label{eq11_15_8}
\frac{1}{c^2}\frac{\pa\varphi}{\pa t}+ \nabla\cdot\mathbf{A}=0
\end{equation}due to the conservation of free charges.
Both the scalar potential $\varphi$ and the vector potential $\mathbf{A}$ have to be continuous across the interfaces of media \cite{1,8,2}. It has been shown in \cite{2} that a sufficient set of  boundary conditions is the continuities of $\varphi, \mathbf{A}$ and $\mathbf{H}_{\parallel}$.

In the following, we find the Casimir energy of a massive vector field in the configuration depicted in Fig. \ref{f1}, where a perfectly conducting ball of radius $a_1$ is in the center of a perfectly conducting spherical shell occupying the region $a_2<r<a_3$. For simplicity, we only consider the scenario where the thickness of the spherical shell is infinite, i.e., $a_3\rightarrow\infty$.

As discussed in \cite{1,2}, the eigenmodes of a massive vector field can be divided into transverse modes with $\nabla\cdot\mathbf{A}=0$ and longitudinal modes with $\nabla\times \mathbf{A}=\mathbf{0}$. The transverse modes can be further divided into type I modes with $\mathbf{E}_r=0$ and type II modes with $\mathbf{B}_r=0$. They are generalizations of the TE and TM modes for a massless vector field. To satisfy all the boundary conditions of a massive vector field, type I transverse modes can be treated alone and they are called TE modes. However, the   type II transverse modes and the longitudinal modes have to be considered together and their linear combinations are called TM modes \cite{2}.

\subsection{TE modes}   TE modes are type I transverse modes.  The continuity of $\mathbf{H}_{\parallel}$ imply that the potentials  vanish identically in the perfectly conducting objects. In the vacuum between the objects, the potentials are given by $\varphi=0$, $A_r=0$ and
\begin{equation*}\begin{split}
\begin{aligned}
A_{\theta}=&m r^{-1}\left(\mathfrak{A}\tilde{j}_{l }(\lambda r)+\mathfrak{B}\tilde{y}_l(\lambda r)\right)  \frac{P_{l}^m(\cos\theta)}{\sin\theta}e^{im\phi}e^{-i\omega t},\\
A_{\phi}=&i r^{-1}\left(\mathfrak{A}\tilde{j}_{l }(\lambda r)+\mathfrak{B}\tilde{y}_l(\lambda r)\right) \frac{dP_{l}^m(\cos\theta)}{d\theta}  e^{im\phi}e^{-i\omega t},\end{aligned}
\end{split}\end{equation*} where  $ l=1,2,\ldots,\;-l\leq m\leq l$,
$$\lambda^2=\frac{\omega^2}{c^2}-\frac{m^2c^2}{\hbar^2},$$  $P_l^m(z)$ are associated Legendre polynomials and  $\tilde{j}_l(z)$ and $\tilde{y}_l(z)$ are the   Riccati-Bessel functions   defined by
$$\tilde{j}_l(z)=\sqrt{\frac{\pi z}{2}}J_{l+\frac{1}{2}}(z),\hspace{1cm}\tilde{y}_l(z)=\sqrt{\frac{\pi z}{2 }}Y_{l+\frac{1}{2}}(z).$$$J_{\nu}(z)$ and $Y_{\nu}(z)$ are the Bessel functions of first kind and second kind.
The continuity of $\mathbf{A}_{\parallel}$ gives the boundary conditions:
\begin{equation*}
\begin{split}
\mathfrak{A}\tilde{j}_{l }(\lambda a_1)+\mathfrak{B}\tilde{y}_l(\lambda a_1)=0,\\
\mathfrak{A}\tilde{j}_{l }(\lambda a_2)+\mathfrak{B}\tilde{y}_l(\lambda a_2)=0.
\end{split}\end{equation*}Therefore, the TE eigenfrequencies are solutions of the equation
$$\tilde{j}_{l }(\lambda a_2)\tilde{y}_l(\lambda a_1)-\tilde{j}_{l }(\lambda a_1) \tilde{y}_l(\lambda a_2)=0.$$
The TE modes are direct generalizations of the TE modes in the massless case \cite{3,4,6,5,12}.  By requiring that the Casimir interaction energy between the spherical objects to be renormalized so that it approaches zero when the separation between the objects becomes infinite, one finds immediately as in \cite{12,2} that the TE contribution to the renormalized  Casimir interaction energy is given by
\begin{equation*}
\begin{split}
E_{\text{Cas}}^{\text{TE}} =&  \frac{\hbar}{2\pi}\sum_{l=1}^{\infty}(2l+1)\int_0^{\infty}\ln \Delta^l_{\text{TE}}(i\xi)d\xi,
\end{split}
\end{equation*}where
\begin{equation*}
\Delta^l_{\text{TE}}(i\xi)= 1-\frac{s_l(\gamma a_1)e_l(\gamma a_2)}{e_l(\gamma a_1)s_l(\gamma a_2)},\quad \gamma:=\sqrt{\frac{\xi^2}{c^2}+\frac{m^2c^2}{\hbar^2}}.
\end{equation*}Here $s_l(z)$ and $e_l(z)$ are modified Riccati-Bessel functions of first kind and second kind defined by
\begin{equation*}
s_l(z)=\sqrt{\frac{\pi z}{2}}I_{l+\frac{1}{2}}(z),\quad e_l(z)=\sqrt{\frac{2z}{\pi}}K_{l+\frac{1}{2}}(z).
\end{equation*}$I_{\nu}(z)$ and $K_{\nu}(z)$ are the modified Bessel functions of first  kind and second kind. Using the fact that $e_l(z)/s_l(z)$ is a positive decreasing function of $z$, we find that the TE contribution to the Casimir interaction force between the two bodies is always attractive.

\subsection{TM modes}  The TM modes are linear combinations of the type II transverse modes and the longitudinal modes. The continuity of $\mathbf{H}_{\parallel}$ implies that in the perfectly conducting bodies, the type II transverse modes have to vanish. Only the longitudinal modes can exist in the perfectly conducting bodies.
Let $$\lambda_0=\frac{\omega}{c},\quad \lambda=\sqrt{\frac{\omega^2}{c^2}-\frac{m^2c^2}{\hbar^2}}.$$
Inside the region  $r<a_1$ occupied by the perfectly conducting ball and the region $a_2<r<a_3$ occupied by the perfectly conducting spherical shell, the potentials of the TM modes are given by
\begin{equation*}\begin{split}
\begin{aligned}
\varphi=&i  \omega r^{-1}\left[\mathfrak{E}_i\tilde{j}_{l }(\lambda_0 r)+\mathfrak{F}_i\tilde{y}_l(\lambda_0 r)\right] P_{l}^m(\cos\theta)e^{im\phi}e^{-i\omega t}\\
A_r=&-r^{-2}\left(\mathfrak{E}_i \left[
 \tilde{j}_{l }(\lambda_0 r)- \lambda_0 r   \tilde{j}_{l }'(\lambda_0 r)  \right]+\mathfrak{F}_i \left[
 \tilde{y}_{l }(\lambda_0 r)- \lambda_0 r   \tilde{y}_{l }'(\lambda_0 r)  \right]\right)
P_{l}^m(\cos\theta)e^{im\phi}e^{-i\omega t}\\
A_{\theta}=& r^{-2}\left[\mathfrak{E}_i\tilde{j}_{l }(\lambda_0 r)+\mathfrak{F}_i\tilde{y}_l(\lambda_0 r)\right] \frac{dP_{l}^m(\cos\theta)}{d\theta}e^{im\phi}e^{-i\omega t}\\
A_{\phi}=& im r^{-2   } \left[\mathfrak{E}_i\tilde{j}_{l }(\lambda_0 r)+\mathfrak{F}_i\tilde{y}_l(\lambda_0 r)\right]
\frac{P_{l}^m(\cos\theta)}{\sin\theta}e^{im\phi}e^{-i\omega t}\end{aligned}.
\end{split}\end{equation*}Here $i=1$ for the region $r<a_1$ and $i=3$ for the region $a_2<r<a_3$. Moreover, $\mathfrak{F}_1=0$. One can check immediately that the electric field and the magnetic field vanish identically in the perfectly conducting bodies.

In the vacuum region $a_1<r<a_2$, the potentials are linear combinations of the type II transverse modes and the longitudinal modes given by
\begin{equation*}
\begin{split}
\begin{aligned}
\varphi=&\frac{ic^2\lambda^2}{\omega}r^{-1}\left[\mathfrak{E}_2\tilde{j}_{l }(\lambda r)+\mathfrak{F}_2\tilde{y}_l(\lambda r)\right] P_{l}^m(\cos\theta)e^{im\phi}e^{-i\omega t}\\
A_r=&r^{-2}\left\{ l(l+1)\left[\mathfrak{C}\tilde{j}_l(\lambda r)+\mathfrak{D}\tilde{y}_l(\lambda r)\right]-  \mathfrak{E}_2\left[\tilde{j}_l(\lambda r)-\lambda r\tilde{j}_l'(\lambda r)\right]-\mathfrak{F}_2\left[\tilde{y}_l(\lambda r)-\lambda r\tilde{y}_l'(\lambda r) \right]\right\}P_l^m(\cos\theta)e^{im\phi}e^{-i\omega t}\\
A_{\theta}=&r^{-2}\left\{ \lambda r \left[ \mathfrak{C} \tilde{j}_l'(\lambda r) +  \mathfrak{D} \tilde{y}_l'(\lambda r) \right] +    \left[\mathfrak{E}_2\tilde{j}_l(\lambda r)+\mathfrak{F}_2\tilde{y}_l(\lambda r)\right]\right\} \frac{dP_l^m(\cos\theta)}{d\theta}e^{im\phi}e^{-i\omega t}\\
A_{\phi}=&im r^{-2}\left\{\lambda r \left[ \mathfrak{C} \tilde{j}_l'(\lambda r) +  \mathfrak{D} \tilde{y}_l'(\lambda r) \right] +     \left[\mathfrak{E}\tilde{j}_l(\lambda r)+\mathfrak{F}\tilde{y}_l(\lambda r)\right]\right\} \frac{ P_l^m(\cos\theta)}{\sin\theta}e^{im\phi}e^{-i\omega t}
\end{aligned}.\end{split}
\end{equation*}
The continuities of $\varphi, A_r,\mathbf{A}_{\parallel}$ give rise to the following boundary conditions:
 \begin{equation}\label{eq12_9_1}
\begin{split}
\text{1.}\quad &\mathfrak{E}_3\tilde{j}_{l }(\lambda_0 a_3)+\mathfrak{F}_3\tilde{y}_l(\lambda_0 a_3)=0\\
\text{2.}\quad &  \omega  \mathfrak{E}_1\tilde{j}_{l }(\lambda_0 a_1)=\frac{c^2 \lambda^2}{\omega}\left[\mathfrak{E}_2\tilde{j}_{l }(\lambda a_1)+\mathfrak{F}_2\tilde{y}_l(\lambda a_1)\right]\\
\text{3.}\quad &\omega\left(\mathfrak{E}_3\tilde{j}_{l }(\lambda_0 a_2)+\mathfrak{F}_3\tilde{y}_l(\lambda_0 a_2)\right)=\frac{c^2 \lambda^2}{\omega}\left(\mathfrak{E}_2\tilde{j}_{l }(\lambda a_2)+\mathfrak{F}_2\tilde{y}_l(\lambda a_2)\right)\\
\text{4.}\quad &-\left(\tilde{j}_{l }(\lambda_0 a_1)-\lambda_0  a_1 \tilde{j}_{l }'(\lambda_0 a_1)\right) \mathfrak{E}_1  \\&=  l(l+1)\left[ \mathfrak{C}\tilde{j}_l(\lambda a_1)+\mathfrak{D}\tilde{y}_l(\lambda a_1)\right] -  \left(\tilde{j}_l(\lambda a_1)-\lambda a_1\tilde{j}_l'(\lambda a_1)\right)\mathfrak{E}_2-\left(\tilde{y}_l(\lambda a_1)-\lambda a_1\tilde{y}_l'(\lambda a_1)\right)\mathfrak{F}_2\\
\text{5.}\quad &-\left(\tilde{j}_{l }(\lambda_0 a_2)-\lambda_0  a_2 \tilde{j}_{l }'(\lambda_0 a_2)\right) \mathfrak{E}_3
 -\left(\tilde{y}_{l }(\lambda_0 a_2)-\lambda_0  a_2 \tilde{y}_{l }'(\lambda_0 a_2)\right) \mathfrak{F}_3\\ &=  l(l+1)\left[ \mathfrak{C}\tilde{j}_l(\lambda a_2)+\mathfrak{D}\tilde{y}_l(\lambda a_2) \right]-  \left(\tilde{j}_l(\lambda a_2)-\lambda a_2\tilde{j}_l'(\lambda a_2)\right)\mathfrak{E}_2-\left(\tilde{y}_l(\lambda a_2)-\lambda a_2\tilde{y}_l'(\lambda a_2)\right)\mathfrak{F}_2\hspace{2cm}\\
\text{6.} \quad& \mathfrak{E}_1\tilde{j}_{l }(\lambda_0 a_1)  = \lambda a_1 \left[ \mathfrak{C} \tilde{j}_l'(\lambda a_1) + \mathfrak{D}\tilde{y}_l'(\lambda a_1) \right] +     \left[\mathfrak{E}_2\tilde{j}_l(\lambda a_1)+\mathfrak{F}_2\tilde{y}_l(\lambda a_1)\right]\\
\text{7.}\quad &\mathfrak{E}_3\tilde{j}_{l }(\lambda_0 a_2)+\mathfrak{F}_3\tilde{y}_l(\lambda_0 a_2)= \lambda a_2   \left[ \mathfrak{C} \tilde{j}_l'(\lambda a_2) + \mathfrak{D}\tilde{y}_l'(\lambda a_2) \right] +     \left[\mathfrak{E}_2\tilde{j}_l(\lambda a_2)+\mathfrak{F}_2\tilde{y}_l(\lambda a_2)\right]
\end{split}
\end{equation}
The eigenfrequencies of the TM modes are those $\omega$ such that this   system of linear equations in $(\mathfrak{C},\mathfrak{D},\mathfrak{E}_1,\mathfrak{E}_2,\mathfrak{E}_3,\mathfrak{F}_2,\mathfrak{F}_3)$ has a nontrivial solution. Using the same techniques as in \cite{13,2}, one  finds   that the TM contribution to the renormalized  Casimir interaction energy is given by
\begin{equation*}
\begin{split}
E_{\text{Cas}}^{\text{TM}}(s)=&  \frac{\hbar}{2\pi}\sum_{l=1}^{\infty}(2l+1)\int_0^{\infty}\ln \Delta^l_{\text{TM}}(i\xi)d\xi,
\end{split}
\end{equation*}where
\begin{equation*}
\Delta^l_{\text{TM}}(i\xi)=\frac{\det Q^l(\xi)}{\det Q^{l}_0(\xi)},
\end{equation*}$Q^l=(Q^l_{ij})$  is the $4\times 4$ matrix
$$Q^l(\xi)=\begin{pmatrix} W_1^l & W_2^l\\ W_3^l & W_4^l\end{pmatrix}$$with
\begin{equation*}
\begin{split}
W_1^l=&\begin{pmatrix} Q_{11}^l & Q_{12}^l\\Q^l_{21} & Q^l_{22}\end{pmatrix}=\left(\begin{aligned} \gamma a_1s_l'(\gamma a_1) \quad &  \gamma a_1e_l'(\gamma a_1)\\
 \gamma a_2s_l'(\gamma a_2)\quad &  \gamma a_2e_l'(\gamma a_2) \end{aligned}\right),\\
W_2^l=&\begin{pmatrix} Q^l_{13} & Q^l_{14}\\Q^l_{23} & Q^l_{24}\end{pmatrix}=\left(\begin{aligned}
- \frac{m^2c^2}{\hbar^2}s_l(\gamma a_1)\quad &  -\frac{m^2c^2}{\hbar^2}e_l(\gamma a_1)\\
 -\frac{m^2c^2}{\hbar^2}s_l(\gamma a_2) \quad & -\frac{m^2c^2}{\hbar^2}e_l(\gamma a_2)\end{aligned}\right),
\\
W_3^l=&\begin{pmatrix} Q^l_{31} & Q^l_{32}\\Q^l_{41} & Q^l_{42}\end{pmatrix}=\left(\begin{aligned} l(l+1) s_l(\gamma_0a_1)s_l(\gamma a_1)\quad &   l(l+1) s_l(\gamma_0a_1)e_l(\gamma a_1)\\
l(l+1)e_l(\gamma_0a_2)s_l(\gamma a_2)\quad &  l(l+1)e_l(\gamma_0a_2)e_l(\gamma a_2)\end{aligned}\right),\\
W_4^l=&\begin{pmatrix} Q^l_{33} & Q^l_{34}\\Q^l_{43} & Q^l_{44}\end{pmatrix}=\left(\begin{aligned} \gamma^2s_l(\gamma a_1)\tilde{s}_l(\gamma_0 a_1)-\gamma_0^2s_l(\gamma_0 a_1)\tilde{s}_l(\gamma  a_1)
\quad & \gamma^2e_l(\gamma a_1)\tilde{s}_l(\gamma_0 a_1) -\gamma_0^2s_l(\gamma_0 a_1)\tilde{e}_l(\gamma  a_1) \\
\gamma^2s_l(\gamma a_2)\tilde{e}_l(\gamma_0 a_2)
-\gamma_0^2e_l(\gamma_0 a_2)\tilde{s}_l(\gamma  a_2) \quad & \gamma^2e_l(\gamma a_2)\tilde{e}_l(\gamma_0 a_2)
-\gamma_0^2e_l(\gamma_0 a_2)\tilde{e}_l(\gamma  a_2) \end{aligned}\right),
\end{split}
\end{equation*}and $Q_0^l$ is the $4\times 4$ matrix
$$Q_0^l=\begin{pmatrix} Q_{11}^l & Q_{12}^l & Q^l_{13} & Q^l_{14}\\Q^l_{21} & 0 & Q^l_{23} & 0\\Q^l_{31} & Q^l_{32} & Q^l_{33} & Q^l_{34} \\
Q^l_{41} & 0 & Q^l_{43} & 0\end{pmatrix}=\left(Q^l_{12}Q^l_{34}-Q^l_{32}Q^l_{14}\right)\left(Q^l_{21}Q^l_{43}-Q^l_{23}Q^l_{41}\right).$$
Here
$$\gamma=\sqrt{\frac{\xi^2}{c^2}+\frac{m^2c^2}{\hbar^2}},\quad \gamma_0=\frac{\xi}{c},$$
$$\tilde{s}_l(z)=s_l(z)-zs_l'(z),\quad \tilde{e}_l(z)=e_l(z)-ze_l'(z).$$ The expansions of $\det Q^l(\xi)$ and $\det Q_0^l(\xi)$ are given in the Appendix.

In the massless limit $m= 0$, $\gamma=\gamma_0$. Therefore,  $W_2^l=0$,
$$W_4^l=\gamma_0^3\begin{pmatrix}0 &-a_1\\ a_2 & 0\end{pmatrix},$$and we find that
$$\Delta_l^{\text{TM}}(i\xi)=1-\frac{s_l'(\gamma_0a_1)e_l'(\gamma_0a_2)}{e_l'(\gamma_0a_1)s_l'(\gamma_0a_2)}.$$This is precisely  the TM contribution to the Casimir interaction energy between two perfectly conducting spherical shells due to a massless vector field \cite{3,4,6,5,12}.

\section{Discussion and Conclusion}

In the case of two parallel perfectly conducting plates, the TM modes can be decomposed into discrete modes whose potentials vanish identically in the perfectly conducting plates and continuum modes whose potentials do not vanish in the perfectly conducting plates. However, for perfectly conducting spherical bodies, one cannot find TM modes that vanish identically inside the perfectly conducting bodies.  In fact, one can check that if one impose the conditions that $\mathfrak{E}_1=\mathfrak{E}_3=\mathfrak{F}_3=0$ so that the potentials vanish identically on the perfectly conducting bodies, the boundary conditions \eqref{eq12_9_1} imply that $\mathfrak{C}=\mathfrak{D}=\mathfrak{E}_2=\mathfrak{F}_2=0$, which means that the potentials are identically zero. This can be considered as a major difference between planar and nonplanar objects for a massive vector field. \begin{figure}[h]
\epsfxsize=0.49\linewidth \epsffile{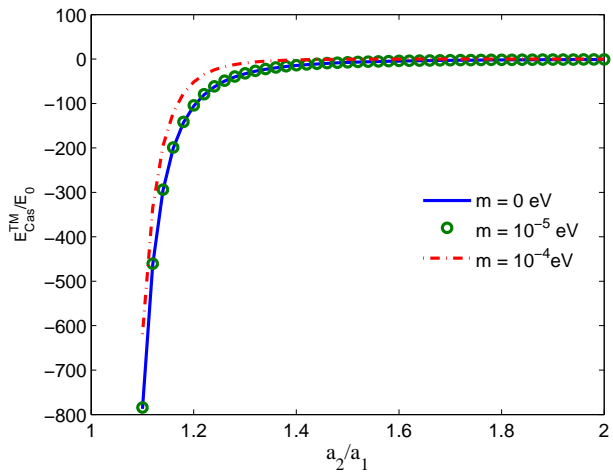} \caption{\label{f2} The TM contribution to the Casimir interaction energy as a function of $a_2/a_1$ when $a_1=1$cm. Here $\displaystyle E_0= \hbar c/(2\pi a_1)$ so that $E_{\text{Cas}}^{TM}/E_0$ is dimensionless. }\end{figure}

\begin{figure}[h]
\epsfxsize=0.49\linewidth \epsffile{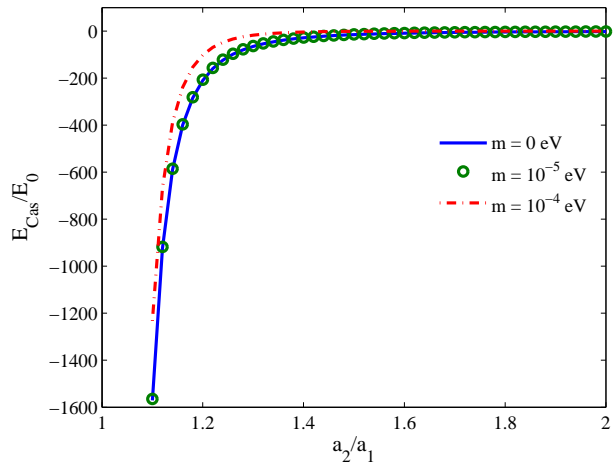} \caption{\label{f3} The total   Casimir interaction energy as a function of $a_2/a_1$ when $a_1=1$cm.  }\end{figure}

\begin{figure}[h]
\epsfxsize=0.49\linewidth \epsffile{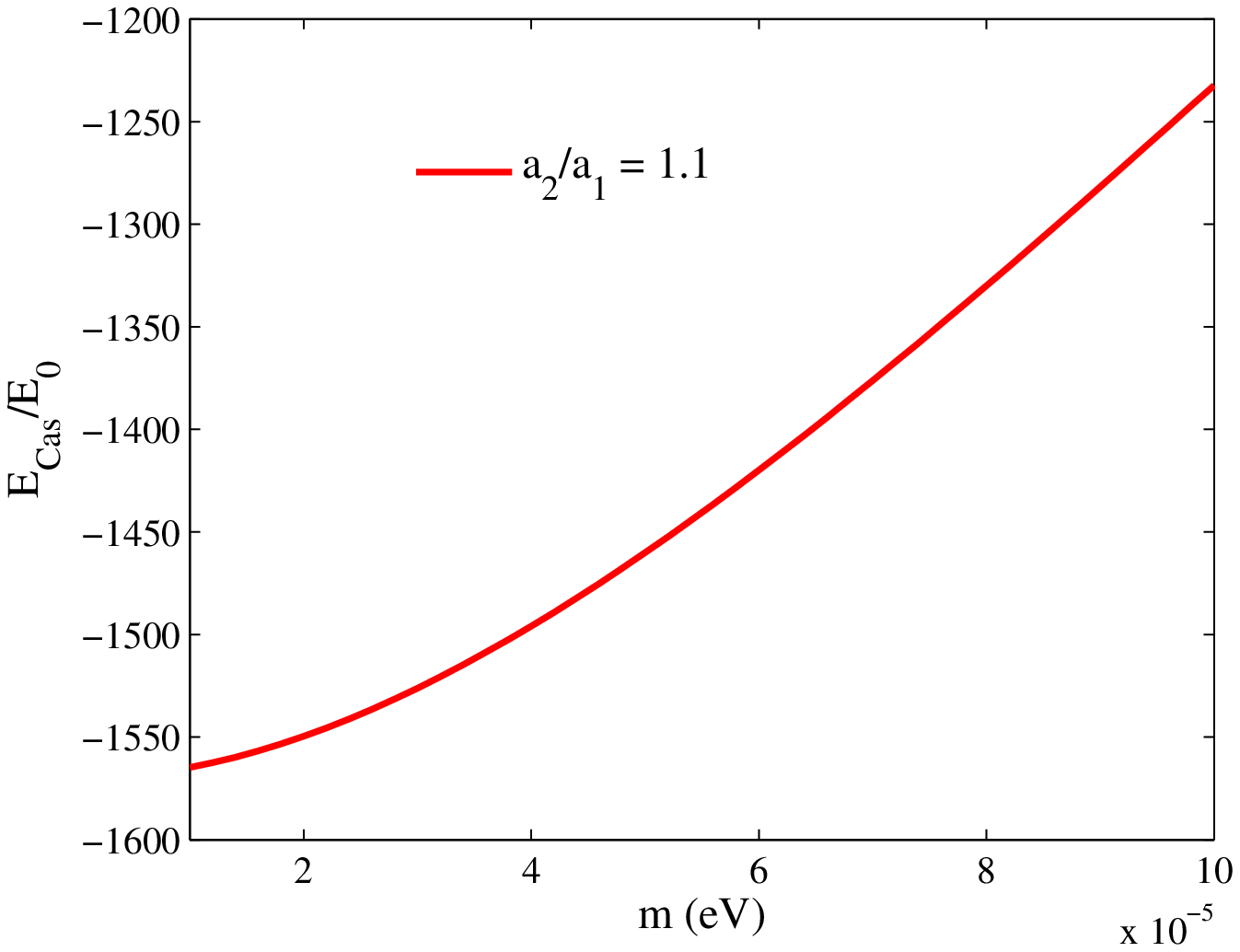} \epsfxsize=0.49\linewidth \epsffile{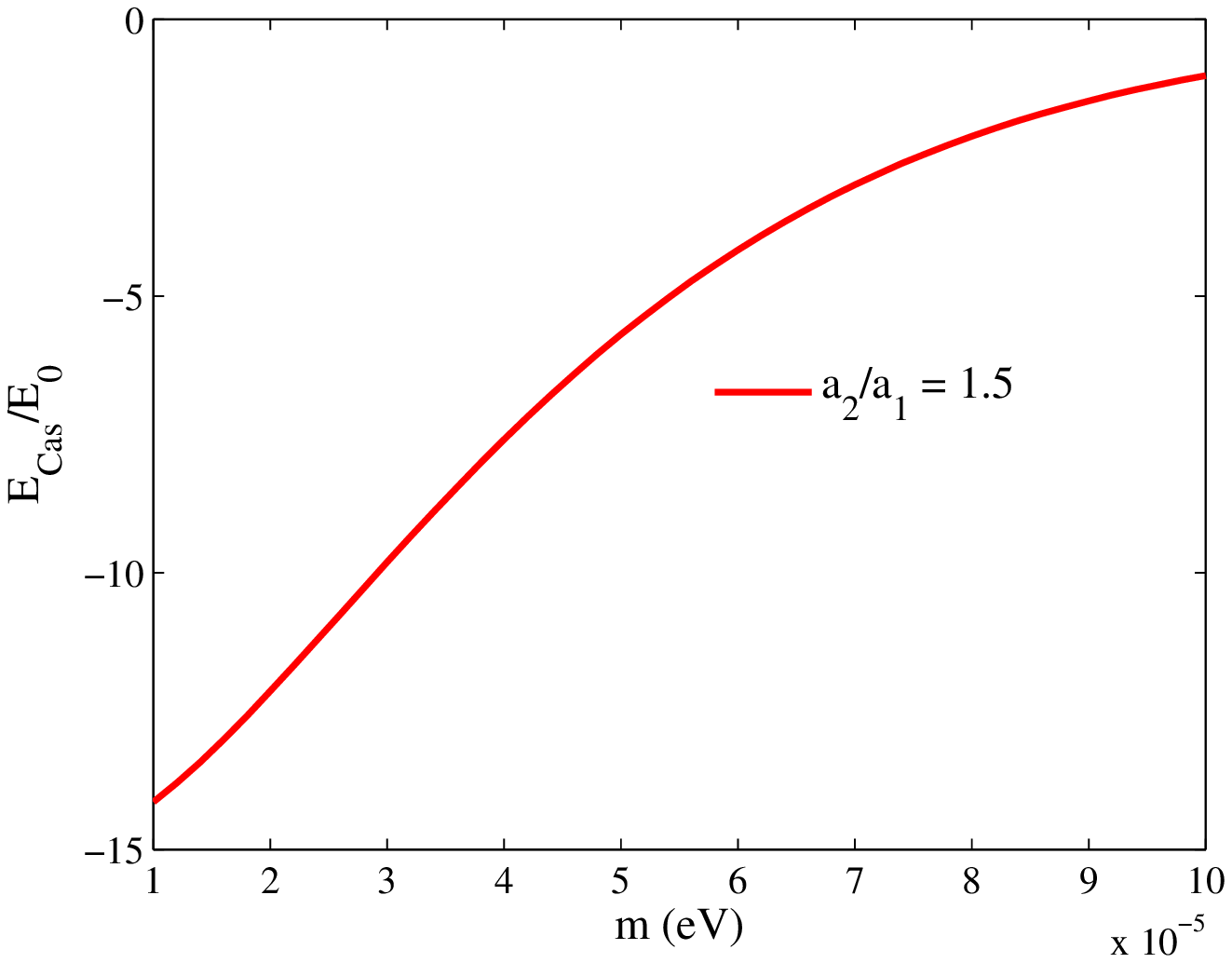}\caption{\label{f4} The total   Casimir interaction energy as a function of mass when $a_1=1$cm and $a_2/a_1=1.1$ or $1.5$.  }\end{figure}

In Fig. \ref{f2}, we plot the dependence of the TM contribution to the Casimir interaction energy as a function of $a_2/a_1$. Here we choose $a_1=1$cm, and we consider the cases where $m=0, m=10^{-5}$eV and $m=10^{-4}$eV. In all these cases, we find from the graph that the TM contribution to the Casimir interaction energy is always attractive. The total Casimir interaction energy as a function of $a_2/a_1$ is plotted in Fig. \ref{f3}. Although the analytical expressions for the TE and TM contributions to the Casimir interaction energy are very different,   TE and TM modes contribute about the same to the total Casimir interaction energy. This is analogous to the case of two parallel perfectly conducting plates. In Fig. \ref{f4}, we show the dependence of the total Casimir interaction energy on mass when $a_2/a_1=1.1$ and $a_2/a_1=1.5$. It is observed that mass corrections can significantly change the magnitude of the Casimir energy. In fact it can be verified analytically that the Casimir interaction energy tends to zero when the mass tends to infinity. 

In conclusion, we have studied the Casimir effect on two perfectly conducting spherical bodies  due to a massive vector field. In contrast to the scenario of two perfectly conducting plates, there are no discrete TM modes that vanish identically in the perfectly conducting spherical bodies. As a result, the analytical formula for the TM contribution to the Casimir energy is much more complicated. Nevertheless, numerical simulations show that the Casimir interaction force is  attractive.

\appendix \section{Alternative expression for $\Delta^l_{\text{TM}}$}  The expansions of the determinants  $\det Q(\xi)$ and $\det Q_0^l(\xi)$ that appear in $\Delta^l_{\text{TM}}(i\xi)$ are given by
\begin{equation*}\begin{split}
\det Q^l(\xi)=& \mathcal{A}^l(\xi)+ \frac{m^2c^2}{\hbar^2}l(l+1)\mathcal{B}^l(\xi)+\left(\frac{m^2c^2}{\hbar^2}\right)^2l^2(l+1)^2\mathcal{C}^l(\xi),\\
\det Q_0^l(\xi)=& \mathcal{A}_0^l(\xi)+  \frac{m^2c^2}{\hbar^2}l(l+1)\mathcal{B}_0^l(\xi)+\left(\frac{m^2c^2}{\hbar^2}\right)^2l^2(l+1)^2\mathcal{C}_0^l(\xi),\end{split}
\end{equation*}where
\begin{equation*}
\begin{split}
\mathcal{A}^l(\xi)=&\Bigl(\bigl[\gamma^2 e_l(\gamma a_1)\tilde{s}_l(\gamma_0 a_1) - \gamma_0^2s_l(\gamma_0 a_1)\tilde{e}_l(\gamma  a_1)\bigr]\bigl[\gamma^2 s_l(\gamma a_2)\tilde{e}_l(\gamma_0 a_2)-\gamma_0^2 e_l(\gamma_0 a_2)\tilde{s}_l(\gamma  a_2)\bigr] \\
 & -\bigl[ \gamma^2 s_l(\gamma a_1)\tilde{s}_l(\gamma_0 a_1)-\gamma_0^2 s_l(\gamma_0 a_1)\tilde{s}_l(\gamma  a_1)\bigr]
 \bigl[ \gamma^2e_l(\gamma a_2)\tilde{e}_l(\gamma_0 a_2)- \gamma_0^2e_l(\gamma_0 a_2)\tilde{e}_l(\gamma  a_2)\bigr]\Bigr)\\&\times  \Bigl(\gamma a_1e_l'(\gamma a_1)\gamma a_2s_l'(\gamma a_2)- \gamma a_1s_l'(\gamma a_1)\gamma a_2e_l'(\gamma a_2)\Bigr),\\
\mathcal{B}^l(\xi)=&s_l(\gamma_0a_1)\Bigl(\gamma a_2s_l'(\gamma a_2)  e_l(\gamma a_1) -\gamma a_2e_l'(\gamma a_2) s_l(\gamma a_1)   \Bigr)\\
&\times\Bigl(e_l(\gamma a_1)\left[ \gamma^2s_l(\gamma a_2)\tilde{e}_l(\gamma_0 a_2)-\gamma_0^2 e_l(\gamma_0 a_2)\tilde{s}_l(\gamma  a_2)\right]-s_l(\gamma a_1)\left[ \gamma^2e_l(\gamma a_2)\tilde{e}_l(\gamma_0 a_2)-\gamma_0^2 e_l(\gamma_0 a_2)\tilde{e}_l(\gamma  a_2)\right]\Bigr)\\
&+e_l(\gamma_0a_2)\Bigl(\gamma a_1s_l'(\gamma a_1) e_l(\gamma a_2)
-\gamma a_1e_l'(\gamma a_1)   s_l(\gamma a_2) \Bigr)\\&\times\Bigl(e_l(\gamma a_2)
\left[ \gamma^2 s_l(\gamma a_1)\tilde{s}_l(\gamma_0 a_1)-\gamma_0^2 s_l(\gamma_0 a_1)\tilde{s}_l(\gamma  a_1)\right]-s_l(\gamma a_2)\left[ \gamma^2e_l(\gamma a_1)\tilde{s}_l(\gamma_0 a_1) -\gamma_0^2 s_l(\gamma_0 a_1)\tilde{e}_l(\gamma  a_1)\right]\Bigr)\\&-2  \gamma^2 \gamma_0^2a_1a_2s_l(\gamma_0 a_1)e_l(\gamma_0 a_2),\\
\mathcal{C}^l(\xi)=&s_l(\gamma_0 a_1)e_l(\gamma_0 a_2)\Bigl(e_l(\gamma a_1)s_l(\gamma a_2)-s_l(\gamma a_1)e_l(\gamma a_2)\Bigr)^2,\hspace{7cm}
\end{split}
\end{equation*}and
\begin{equation*}
\begin{split}
\mathcal{A}_0^l(\xi)=&\gamma a_1e_l'(\gamma a_1)\gamma a_2s_l'(\gamma a_2)\bigl[\gamma^2 e_l(\gamma a_1)\tilde{s}_l(\gamma_0 a_1) - \gamma_0^2s_l(\gamma_0 a_1)\tilde{e}_l(\gamma  a_1)\bigr]\bigl[\gamma^2 s_l(\gamma a_2)\tilde{e}_l(\gamma_0 a_2)-\gamma_0^2 e_l(\gamma_0 a_2)\tilde{s}_l(\gamma  a_2)\bigr],\\
\mathcal{B}_0^l(\xi)=&s_l(\gamma a_2)^2\gamma a_1e_l'(\gamma a_1)e_l(\gamma_0a_2)\bigl[\gamma^2 e_l(\gamma a_1)\tilde{s}_l(\gamma_0 a_1) - \gamma_0^2s_l(\gamma_0 a_1)\tilde{e}_l(\gamma  a_1)\bigr]\\
&+e_l(\gamma a_1)^2\gamma a_2s_l'(\gamma a_2)s_l(\gamma_0 a_1)\bigl[\gamma^2 s_l(\gamma a_2)\tilde{e}_l(\gamma_0 a_2)-\gamma_0^2 e_l(\gamma_0 a_2)\tilde{s}_l(\gamma  a_2)\bigr],\\
\mathcal{C}_0^l(\xi)=&s_l(\gamma a_2)^2e_l(\gamma a_1)^2s_l(\gamma_0 a_1)e_l(\gamma_0a_2).
\end{split}
\end{equation*}One can show that $\mathcal{A}^l(\xi), \mathcal{B}^l(\xi),\mathcal{C}^l(\xi),\mathcal{A}_0^l(\xi),\mathcal{B}_0^l(\xi),\mathcal{C}_0^l(\xi)$ are all positive functions of $\xi$.

\end{document}